\documentclass[a4paper]{article} % For LaTeX2e
\usepackage[T1]{fontenc} % add special characters (e.g., umlaute)
\usepackage[utf8]{inputenc} % set utf-8 as default input encoding
\usepackage{ISMIRlbd, cite, times, amsmath, url}
\usepackage{hyperref}
\usepackage{url}
\usepackage{float}
\usepackage{graphicx}
\usepackage{booktabs}
\usepackage{color}

\title{MIDI Miner -- A Python Library for tonal tension and track classification}

\multauthor{Rui Guo$^1$ \hspace{1cm} Dorien Herremans$^2$ \hspace{1cm} Thor Magnusson$^1$} {	$^1$ University of Sussex, United Kingdom\\
	$^2$  Singapore University of Technology and Design, Singapore\\
	{\tt \small r.guo@sussex.ac.uk}%, kat\_agres@ihpc.a-star.edu.sg, \\ \tt\small dorien\_herremans@sutd.edu.sg}
}

\def\authorname{Rui Guo, Dorien Herremans, Thor Magnusson}

\begin{document}

\maketitle

%\begin{abstract}
%The abstract paragraph should be indented 1/2~inch (3~picas) on both left and
%right-hand margins. Use 10~point type, with a vertical spacing of 11~points.
%The word \textsc{Abstract} must be centered, in small caps, and in point size 12. Two
%line spaces precede the abstract. The abstract must be limited to one
%paragraph.

%The text must be confined within a rectangle 152mm  wide and
%252mm ($\sim$9~inches) long. The left margin is 29mm and .
%Use 10~point type with a vertical spacing of 11~points. Times New Roman is the
%preferred typeface throughout. Paragraphs are separated by 1/2~line space, with no indentation.

We present a Python library, called Midi Miner, that can calculate tonal tension and classify different tracks. MIDI (Music Instrument Digital Interface) is a hardware and software standard for communicating musical events between digital music devices. It is often used for tasks such as music representation, communication between devices, and even music generation \cite{herremans2017functional}.  Tension is an essential element of the music listening experience, which can come from a number of musical features including timbre, loudness and harmony \cite{Farbood2012A}. Midi Miner provides a Python implementation for the tonal tension model based on the spiral array \cite{chew2014mathematical} as presented by Herremans and Chew \cite{herremansTensionRibbonsQuantifying}. Midi Miner also performs key estimation and includes a track classifier that can disentangle melody, bass, and harmony tracks. Even though tracks are often separated in MIDI files, the musical function of each track is not always clear. The track classifier keeps the identified tracks and discards messy tracks, which can enable further analysis and training tasks.    

\textbf{The spiral array} \cite{chew2014mathematical} is a three-dimensional mathematical model of tonality developed by Prof. Elaine Chew. It consists of a helix for pitch classes, chords, and keys. It is important to note that this model takes pitch spelling into account, meaning that Ab and G\# will have a different position in the pitch class helix. Since MIDI files do not capture pitch spelling, we include an algorithm for estimating pitch spelling. First, the key index of the MIDI file is set as the most frequently occurring pitch class. 
The mode of the key (major or minor) cannot be differentiated at this stage, but the key index can inform the pitch spelling.  Based on the estimated key index, the piece will fall into one of two classes, the first is based on the key indices C, D, E, G, A, and B, and the second on Db, Eb, F, Gb, Ab ad Bb. The accidentals of pieces in the first class will be mapped to sharps if needed, whereas the accidentals of pieces in the second class will be mapped to flats. 
After the pitch spelling has been estimated for all note, the final key can be determined using the spiral array key detection algorithm \cite{chuan2005polyphonic}. 

%ASKED and the difference between the mean position of all notes of that song in the spiral array and the position of each key can help to decide if it is major or minor. 

\begin{figure}[htb]
\centering
 \includegraphics[width=.9\columnwidth]{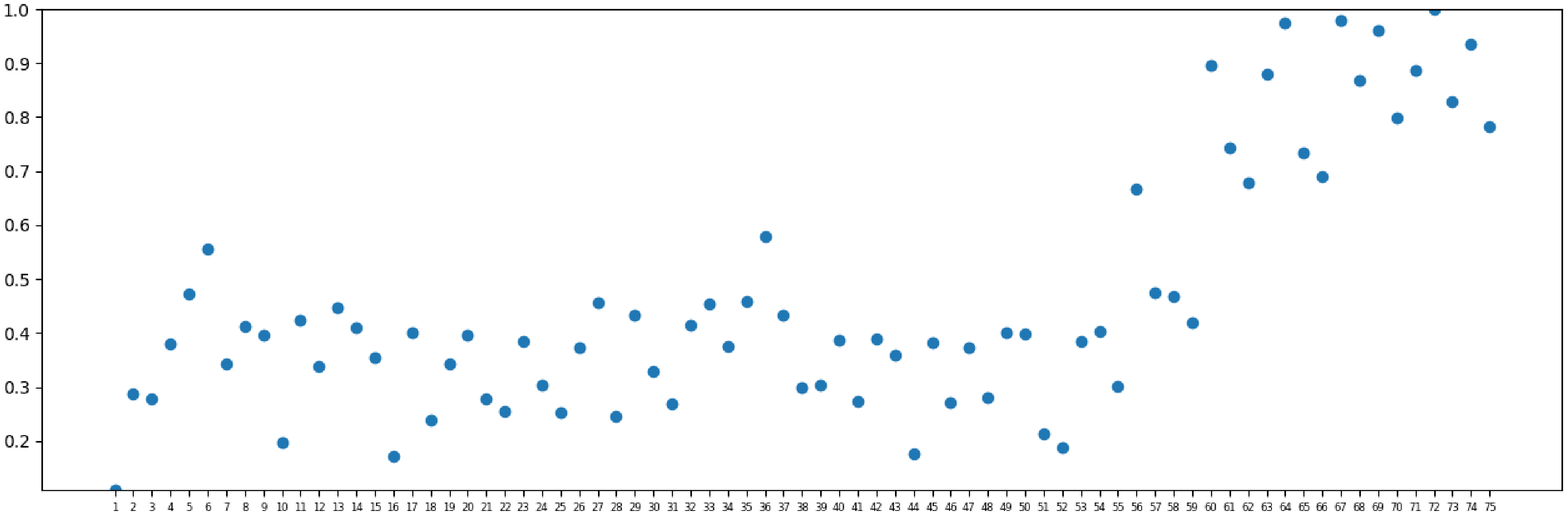}
% \vspace{-.5cm}
\caption[Tensile strain for each bar of the MIDI file available online.]{Tensile strain for each bar of the MIDI file available online\footnote{\url{https://github.com/ruiguo-bio/midi-miner/blob/master/example/new.mid}}.}
 \label{fig:tensile}
\end{figure}

\textbf{Tonal tension}
\cite{herremansTensionRibbonsQuantifying} define three measures of tonal tension: cloud diameter (dissonance), cloud momentu (tonal movement), and tensile strain (distance to the key). Mini Miner implements these three tension measures based on a default window length of one half note (customizable by the user). The implemented model works for polyphonic music, by weighing the occurrence of each pitch class per time window. After inputting a MIDI file, Midi Miner also provides a graphical representation of the three tension measures. Figure \ref{fig:tensile} provides an example output for the tensile strain measure (i.e., distance to the key). A sliding window of 16 beats with average tensile strain is used to detect potential key changes by comparing the ratio between the mean tensile strain and the adjacent window. If this $ratio>2$ for four consecutive windows, then a key change is flagged. In Figure 1, Midi Miner detects a key change in bar 61.

\textbf{Midi track classifier} Midi Miner also implements a track classifier that untangles melody, bass, and harmony tracks. These tracks are of interest for anybody preparing a dataset for, e.g., music generation. A random forest classifier was trained by using 5,006 Chinese pop songs collected from the Internet. Among those files, 3,265 files have at least one track label. These MIDI track labels are strings marked by the author, which usually provide information about the track type. By looking at the labels in those MIDI files, we manually added specific labels for melody, bass, chords, which were used to train our classifier. The drum tracks are currently discarded. Similar to \cite{rizo2006melody}, who select 34 features to classify the melody track, we computed 30 features for each track. These are used as input to train our classifier and are described in detail on our github page. The performance of the track classifier is evaluated in Table \ref{tab:trackresult} using a test set (25\% or the original dataset). Based on the classification result, Midi Miner will extract the melody, bass and harmony tracks and output a new MIDI file with only those tracks.

\begin{table}[hbt]
\begin{center}
\begin{tabular}{llllll}
\toprule
\textbf{track} & \textbf{prediction} & \textbf{precision} & \textbf{recall} & \textbf{F1 score} \\
\midrule
melody & False & 0.99 & 0.99 & 0.99\\ 
& True & 0.95 & 0.93 & 0.94\\ 

bass & False & 0.98 & 1.00 & 0.99\\ 
& True & 0.97 & 0.92 & 0.95\\ 

harmony & False & 0.98 & 0.98 & 0.98\\ 
& True & 0.92 & 0.89 & 0.91\\ 
\bottomrule
\end{tabular}
\caption{Track classification result}
\label{tab:trackresult}
\end{center}
\end{table}

\textbf{Download Midi Miner} The Midi Miner library is available online\footnote{\url{https://github.com/ruiguo-bio/midi-miner}}. In the future, we aim to include a more comprehensive extraction of MIDI features, as well as more complete music analysis tools.

%\end{abstract}

%\section{Final instructions}
%Do not change any aspects of the formatting parameters in the style files.
%In particular, do not modify the width or length of the rectangle the text
%should fit into, and do not change font sizes (except perhaps in the
%\textsc{References} section; see below). Please note that pages should be
%numbered.

%\subsubsection*{Acknowledgments}
\begin{acknowledgments}
This work has received funding from the Chinese scholarship council and MOE T2 grant number MOE2018-T2-2-161.
\end{acknowledgments}

\bibliography{ismir_lbd2019}

\end{document}